# Ultrathin Organic Transistors on Oxide Surfaces[**]

By *Maren Daraktchiev, Adrian von Mühlenen, Frank Nüesch, Michel Schaer,*

*Martin Brinkmann, Marie-Noëlle Bussac, Libero Zuppiroli*[*]

In recent years, thin-film organic field-effect transistors (OFETs) have begun to be considered as a possible alternative to the hydrogenated amorphous silicon thin-film transistors (a-Si:H TFT's) used in active matrix flat panel displays and other large-area electronics applications.[1,2] Low-temperature processability, low-cost fabrication and compatibility with arbitrary substrates are some of the promising advantages of OFETs, among others.[3-5] Of the many organic materials available, pentacene, in particular, is one of the leading candidates for use in current thin-film OFET architectures - this because of its excellent electrical characteristics and its resistance to atmospheric oxygen.[6] In the

[*] Dr. M. Daraktchiev, Mr. A. von Mühlenen, Dr. F. Nüesch, Mr. M. Schaer, Prof. L. Zuppiroli
Swiss Federal Institute of Technology (EPFL)
Institute of Materials (IMX)
CH-1015 Lausanne, Switzerland
e-mail : libero.zuppiroli@epfl.ch

Dr. M. Brinkmann
Institut Charles Sadron
67083 Strasbourg Cedex, France

Prof. Marie-Noëlle Bussac
Centre de Physique Théorique
Ecole Polytechnique
91128 Palaiseau Cedex, France

[**] This work was supported by the Swiss National Science Foundation. We want to thank S. Pratontep for his assistance in AFM observations, Ph. Bugnon for purification of pentacene, J. Kuster for his help in OFETs characterizations, and W. Leo for critically reading the manuscript

recent literature, pentacene's transport properties, as well as transistor performance, have already been analyzed from the point of view of substrate treatments[7,8], pentacene evaporation rate and substrate temperature[9,10], electrode chemical nature and channel geometry[11,12]. The results show that the morphology, crystal structure and molecular ordering of the first organic monolayer(s) at the pentacene/dielectric interface are essential determinants of carrier transport phenomena[9,13-16]. To further investigate these interface effects, we have built a model organic field-effect transistor which consists essentially of a single layer of pentacene on an oxide substrate. Four-probe and two-probe transport measurements as a function of temperature and fields will be presented in relation with structural near-field observations. The experimental results suggest a simple two-dimensional model where the equilibrium between free and trapped carriers at the oxide interface determines the OFET characteristics and performance.

We fabricated a series of thin-film OFETs with different pentacene thicknesses on a $SiO_2$ gate dielectric followed by low-temperature gold deposition of the source-and-drain contacts. The pentacene film thicknesses in these devices ranged from 5-nm (ultrathin) to 100-nm. Also, before each pentacene deposition, the $SiO_2$ surface was activated by exposure to oxygen-plasma for 5 min in a 0.1 mbar $O_2$ atmosphere at a bias of - 40 V.

As will be seen in this paper, measurements on these devices indicate that conduction in the ultrathin transistor involves one or at most two layers of pentacene, even at low gate fields. In ultrathin transistors, therefore, this process may be considered as essentially two-dimensional. Indeed, in the presence of a gate field larger than 0.5 MV/cm the accumulated charges are fully confined to the first layer. We will also demonstrate further on that the ultrathin film transistors have no contact resistances and can therefore be used as a model system. As well, thanks to these properties, we will be able to validate a very simple two-dimensional transport picture, different from the classical FET models[17,18], which illustrates the crucial role played by oxide surface defects in determining the transistor properties.

Figure 1 presents the measured current-voltage characteristics of our OFET's and illustrates the extreme sensitivity of OFET performance to film thickness. Indeed, a close look at this figure shows that the drain current $I_D$ has its maximal value for the ultrathin pentacene transistor and then, surprisingly, drops, in a rather drastic way, as the pentacene thickness increases. Thus, the thinner the pentacene film, the more efficient the transistor in Fig.1.

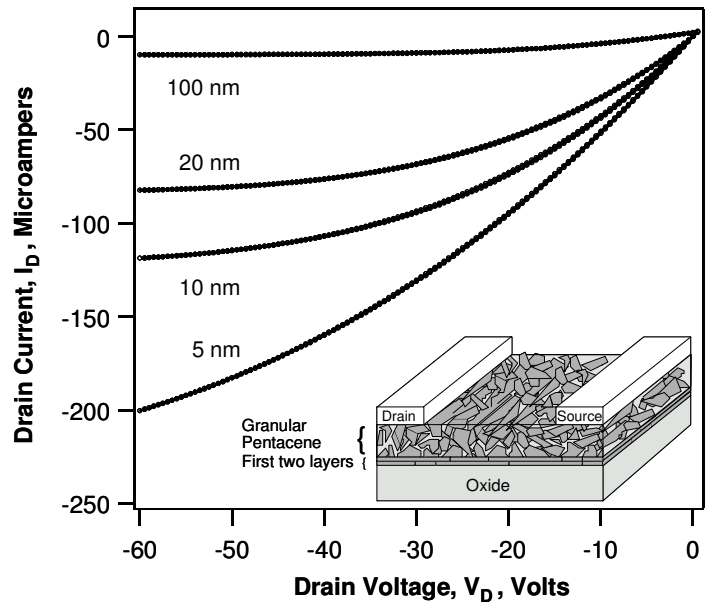

**Figure 1.** Current-voltage characteristics of various pentacene thin film OFETs deposited on oxide surfaces at a gate voltage of - 60 V. The channel length L and width W are 100 and 6000 μm, respectively. Surprisingly, the drain current density decreases as the pentacene thickness increases. The structure of the pentacene film can be depicted as one or two perfectly grown crystalline layers on the gate-oxide dielectric followed by a granular bulk, on top of which a Au drain and source electrode are evaporated.

Let us first discuss this phenomenon from the viewpoint of film growth morphology, which is, in fact, thickness dependent. In Fig.2, the best pentacene morphology is seen to be in the ultrathin transistor where atomic force microscopy images demonstrate full coverage of the adjacent pentacene islands and minimal inter-island boundary density. In particular, to assist the formation of large pentacene grains in the first monolayers on the oxide surface, a substrate temperature and deposition rate of 338 K and 0.6 nm/min, respectively were maintained. Details of the growth conditions may be found in references.[9,10] These values are optimal for the construction of ultrathin pentacene film layers, however, for thicker films, the growth mechanism competes with various coarsening (reconstruction) processes. Thus, inter-island grain boundaries or other crystalline singularities will tend to dominate the bulk of the film. The inset in Fig.1 provides an artist's view of this granular morphology in thick films. Obviously, when the source-drain current flows across the top contacts, the bulk structural imperfections present large potential barriers to the propagation of charge.[19]

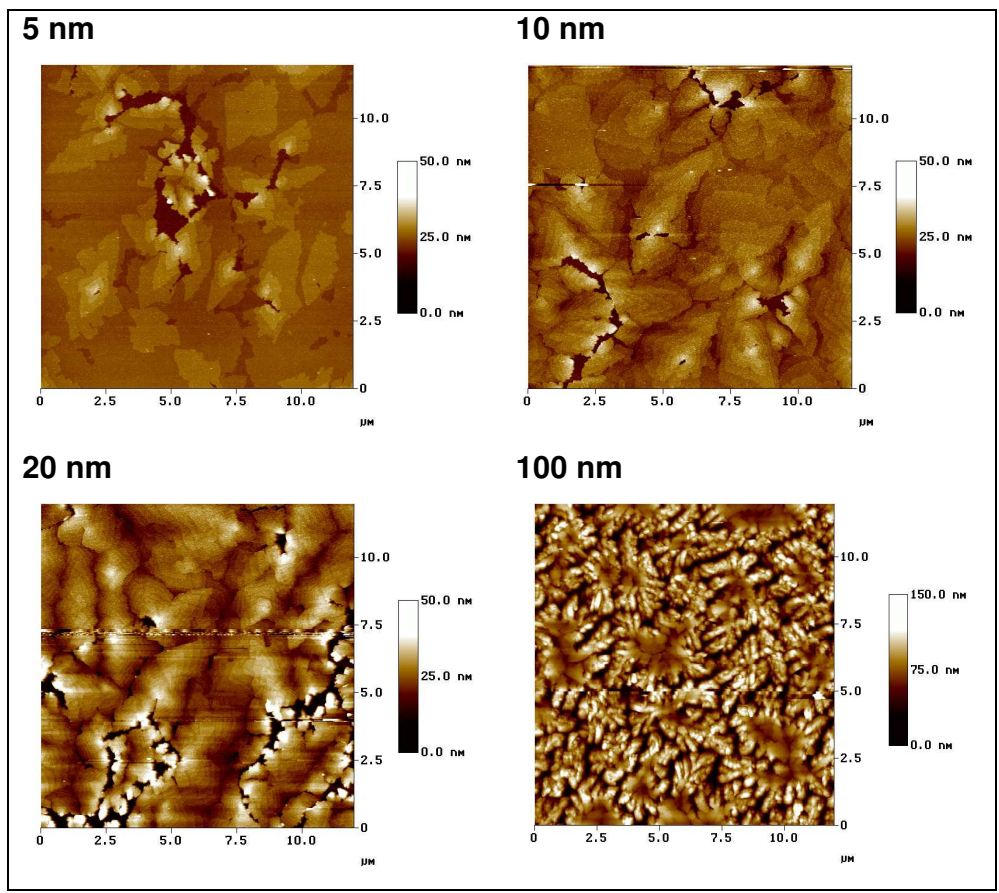

**Figure 2.** AFM images of the conduction channel. The channel morphology changes as the film thickness is increased. The best film structure is in the ultrathin transistor while the worst (dendritic) is in the 100-nm-thick pentacene film where the bulk imperfections determine the film growth process.

In Fig.1, the different electrical characteristics of OFETs at different thicknesses can also be analyzed in terms of the contact resistances of the films. We have investigated two different film-contact geometries: the classical two-probe geometry in which the resistance of the contacts affects the measurements of the intrinsic resistance of the film and the four-probe geometry which eliminates the contact resistance.[18] The results presented in Fig.3 show that the contact resistance is negligible in the ultrathin transistor, while at higher thicknesses it dominates the transistor current.

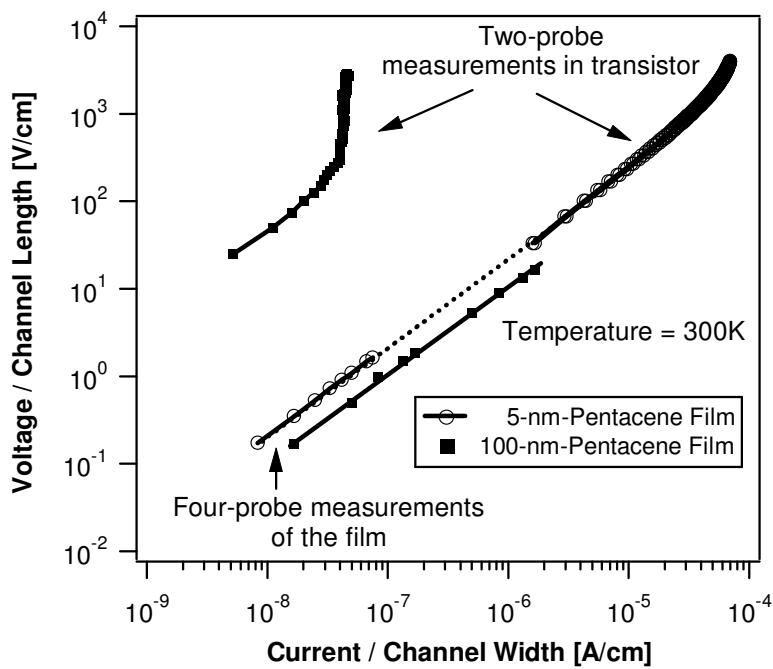

**Figure 3.** Four-probe and two-probe resistivity measurements of the pentacene film. In the ultrathin film transistor the two- and four-probe resistivities coincide indicating that the contact resistance is negligible. In the 100-nm-thick pentacene film OFETs, however, the probes/organic contact resistance becomes a crucial factor for the electrical transport properties. This phenomenon is due to the difficulty of transport from the gold contacts, through the granular bulk, to the semiconducting layer.

Our pentacene films can thus be considered as lamellar structures[20], where only one or two layers close to the oxide are enough to offer a conducting channel for the carriers. This fact is clearly established by the four-probe resistivity measurements on films thicker than 10 nm. Indeed, in Fig.3 the 20-nm- and 100-nm-thick pentacene films exhibit the same two-dimensional resistivity $\rho_{2D}$ at $V_G = 0$. This is fundamentally different from the case of homogeneous films where this transport property should scale inversely with the film thickness. The intrinsic film resistivity $\rho_{2D}$ is found, in fact, to be 35 MΩ/□ which we tend to attribute essentially to the first layer resistance, in agreement with the calculations presented below.

Indeed, when a gate voltage is applied to pentacene OFETs, one can show, in fact, that the above effect is even enhanced. More precisely, the charge density of the first layers, which we denote by i, results from the balance of two forces: the gate voltage electrostatic force that attracts the charge to the layer i=1 closest to the oxide and the entropical force that tends to smear the charge density over further layers according to the microbalance relation

$$\frac{W_{i+1,i}}{W_{i,i+1}} = \exp\frac{-q \cdot (V_{i+1} - V_i)}{k_B T} \quad (1)$$

where q is the carrier charge, T the temperature, $k_B$ the Boltzmann constant and the W's are the transition rates between layers at respective voltage $V_i$. In the continuous limit, the Poisson equation yields the Debye length:[21]

$$\lambda_D = \frac{d \cdot k_B \cdot T}{q \cdot V_G}\sqrt{2} \quad (2)$$

where d is the oxide thickness and $V_G$ is the gate voltage. For a gate field $\frac{V_G}{d}$ greater than 0.5 MV/cm at room temperature ($|V_G|$ = 10 Volts), $\lambda_D$ is less than 0.73 nm. Thus a discrete two-layer model is sufficient to describe the charge densities $\sigma_1$ and $\sigma_2$ within the first and second layers, respectively. From Gauss' theorem, we can deduce the potential on the first layer as

$$V_1 = -V_G + \frac{\sigma_1 + \sigma_2}{\varepsilon_I} = -\frac{\sigma_2}{\varepsilon_S} \qquad (3)$$

where $\varepsilon_I$ and $\varepsilon_S$ are the dielectric permittivities of the oxide and pentacene, respectively.

At thermal equilibrium, $\frac{\sigma_2}{\sigma_1} = \exp\left(\frac{qV_1}{k_BT}\right)$ and $V_1$ is

$$V_1 = -a \cdot \left(\frac{\varepsilon_I}{\varepsilon_S}\right) \frac{V_G}{d + a \cdot \left(\frac{\varepsilon_I}{\varepsilon_S}\right) + d \cdot \exp\left(-\frac{qV_1}{k_BT}\right)} \qquad (4)$$

For $\frac{|V_G|}{d}$ > 0.5 MV/cm, the ratio $\sigma_2/\sigma_1$ is less than 0.15 so that most of the charge is concentrated in the first layer. Note that a discrete n-layer analysis of the charge density would not drastically affect the above results, especially in the high-gate field approach. Thus, conduction in the ultrathin pentacene transistor can be interpreted in terms of a two-dimensional process.

This single active monolayer of pentacene obviously interacts strongly with the oxide surface. However, this interaction is not generally considered in an explicit way in most works concerning OFETs. In some, for example, a somewhat "mysterious" threshold field is introduced at gate fields close to zero to include the main thermodynamic interface effects. Here we propose a different picture, where both the oxide surface and the active pentacene layer are treated as a whole. In this approach, the oxide is not just a homogenous passive dielectric but an active surface in the sense that electroactive surface defects and radicals (peroxy-radicals $\equiv Si-O-O\bullet$, $E'$-centers $\equiv Si\bullet$, nonbridging oxygen hole centers $\equiv Si-O\bullet$ [22-24]) can act as electron acceptors (or hole traps from a pentacene viewpoint). This is particularly true when the surface has been exposed to moisture, or, as in our case, to a short, low energy plasma discharge which activated it prior to the fabrication of the transistor (see Fig.4a).

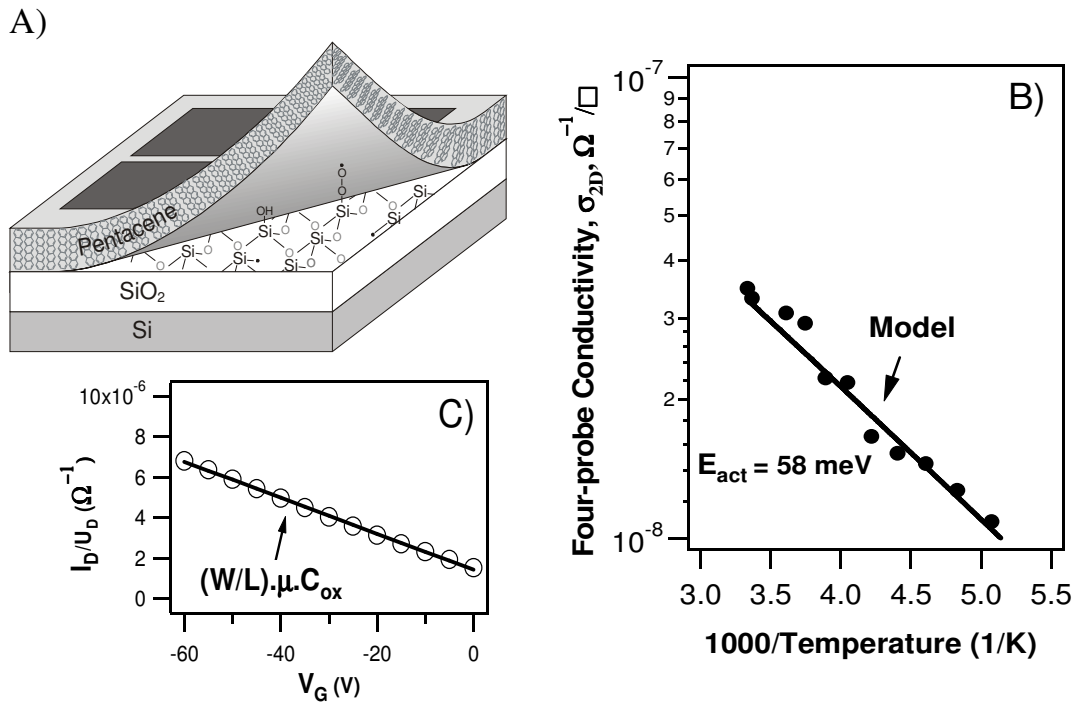

**Figure 4.** **(A)** Sketch of the pentacene/SiO$_2$ interface below the pentacene single layer; **(B)** Intrinsic film two-dimensional conductivity σ$_{2D}$ as a function of temperature. The decline of σ$_{2D}$ with $1/T$ is due to a decrease of the hole density $p_{holes}$ in the first layer; **(C)** The field-effect mobility $\mu$ of the ultrathin OFETs (~ 0.1 cm$^2$/Vs) is calculated from the slope of $\frac{I_D}{V_D}$ versus $V_G$ (transconductance), which itself was derived from the linear regime of the current-voltage transistor characteristics. Here the mobility is independent of the gate voltage due to the absence of contact resistance.

Different types of radicals have been recently identified on oxide (SiO$_2$, TiO$_2$, Al$_2$O$_3$) surfaces.[23,25] One of them is particularly active on SiO$_2$: the peroxy-radicals.[23,24] These oxygen active defects can accept electrons from the pentacene molecules. Similar to electrochemical reactions in a polar liquid phase, the charge transfer reactions at the oxide-pentacene interface are made possible by the solid state charge solvation process. Both oxide and pentacene are polar so that their electronic levels are shifted significantly with respect to the single molecule case.[26]

Experimental proof for the existence of these charge transfer reactions can be found in the four-probe measurements on pentacene films in Fig. 3. These measurements were also repeated systematically for many of the other samples. Obviously, the Ohm law that is observed over several current decades at zero gate field (see the slope of 1 in Fig. 3) indicates the presence of a few ppm of residual holes transferred from the oxide surface into the pentacene lattice. In this approach, very simple equations can be written to describe the interface equilibrium. The chemical potential $\varepsilon_F$ of either electrons trapped on peroxy-radicals or holes transferred to the pentacene molecules can be expressed by the usual Langmuir isotherm on the oxide side[27] and narrow-band statistics on the pentacene side. Thus

$$\varepsilon_F = \Delta\varepsilon - k_B \cdot T \cdot \ln\left(\frac{R}{n} - 1\right) \qquad (5)$$

where $\Delta\varepsilon = \varepsilon_- - E$ is the difference between the electron affinity of the radicals $\varepsilon_-$ and ionization potential E of the pentacene molecules (solvation effects included), R is the radical density on $SiO_2$ per unit area, and n is the density of the electrons ($cm^{-2}$) trapped on the radical levels.

According to Fermi-Dirac statistics, the hole density $p_{holes}$ ($cm^{-2}$) in pentacene is:

$$p_{holes} = \int_{+4J}^{-4J} D(\varepsilon) \cdot f(\varepsilon,\varepsilon_F) \cdot d\varepsilon = \frac{P \cdot k_B \cdot T}{4 \cdot |J|} \cdot \ln\left[\frac{1+\exp\left(-\frac{4 \cdot J + \varepsilon_F}{k_B \cdot T}\right)}{1+\exp\left(\frac{4 \cdot J - \varepsilon_F}{k_B \cdot T}\right)}\right] \qquad (6)$$

where $D(\varepsilon)$ is the two-dimensional density of states per unit area and unit energy, $f(\varepsilon,\varepsilon_F)$ is the Fermi factor of the holes, $J < 0$ is the transfer integral in pentacene, P is the density of pentacene molecules per unit area and $\varepsilon_F$ is the Fermi level of the carriers defined with respect to the center of the pentacene band with width $8 \cdot |J|$.

At thermal equilibrium the two-dimensional charge transfer on the pentacene molecules/$SiO_2$ interface is controlled by the equilibrium of the chemical potentials $\varepsilon_F$ from Eq.5-6. The residual carrier density at $V_G = 0$ satisfies the condition $p_{holes} = n$. Further, when a negative $V_G$ potential is applied to the $SiO_2$ gate dielectric the carrier density in the first layer of OFETs will be the sum of the residual carrier density of the film and the field-effect charge density accumulated on the dielectric semiconductor interface:

$$p_{holes} = -\frac{C_{ox} \cdot V_G}{q} + n = \frac{C_{ox}}{q} \cdot (V_T - V_G), \quad q > 0, \quad V_G < 0 \tag{7}$$

where $C_{ox}$ is the electrical capacity of the gate oxide (19.5 nF/cm$^2$).

The residual carrier concentration n determines the threshold field $V_T = \frac{q \cdot n}{C_{ox}}$ in the ultrathin transistor. Besides, n = $p_{holes}$ also determines the two-dimensional conductivity from the four-probe measurements (see Fig.4b):

$$\sigma_{2D} = p_{holes} \cdot q \cdot \mu \tag{8}$$

where μ is the hole mobility.

From the transconductance characteristics (see Fig.4c) calculated from the linear regime of ultrathin transistor, we deduce both the field effect mobility at room temperature μ = 0.1 cm$^2$/V·s and the threshold voltage $V_T$ = +16 V. Note that the mobility is not gate-voltage dependent because of the absence of contact resistance. The intrinsic conductivity is found to be about 2.86 × 10$^{-8}$ Siemens·□ (see Eq.8) so that the residual concentration of carriers is n (300 K) = $p_{holes}$ (300 K) = 1.88 × 10$^{12}$ cm$^{-2}$. Furthermore, we have neglected the temperature variation of the mobility since this is in agreement with many recent monographs,[28,29] and have attributed the measured conductivity activation energy (~ 58 meV) in Fig.4b to the variation of trapped electron concentrations on the oxide surfaces. Moreover, from Eq.5-6 and the conservation relation of the residual carriers n = $p_{holes}$, the chemical potential $\varepsilon_F(T)$ can be deduced. At the first order approximation (non-degenerated limit), one writes:

$$\varepsilon_F(T) = \frac{\Delta\varepsilon - 4 \cdot J}{2} + k_B \cdot T \cdot \ln\left(\frac{P \cdot k_B \cdot T}{4 \cdot R \cdot |J|}\right) \tag{9}$$

By reporting Eq.9 into Eq.6, we finally obtain the number of residual carriers as a function of temperature

$$p_{holes} = \sqrt{\frac{P \cdot R \cdot k_B \cdot T}{4 \cdot |J|}} \cdot \ln\left[1 + \exp\left(-\frac{4 \cdot J + \Delta\varepsilon}{2 \cdot k_B \cdot T}\right)\right] \tag{10}$$

which can be compared to the experiment in Fig.4b to get $\Delta\varepsilon + 4 \cdot J = 100$ meV. One can also estimate the position of the Fermi level at room temperature $\varepsilon_F = 345$ meV with respect to the center of the pentacene band, the charge transfer integral $|J| = 75$ meV, the charge transfer energy $\Delta\varepsilon = 400$ meV and the radical concentration $R = 5 \times 10^{12}$ cm$^{-2}$.

Despite its extreme simplicity, the transistor model presented here, which is based on only one type of trap (peroxy radicals, for instance) on the dielectric surface shows the importance of describing the charge transfers at the interface. It is particularly successful in determining the pertinent transfer integral $|J|$ in pentacene, the value of which is consistent with the renormalization theory[26] and quantum chemistry calculations.[20] Moreover, the positive threshold field in the ultrathin transistor discussed in this paper can be entirely attributed to the residual carrier concentration through $V_T = \frac{n \cdot q}{C_{ox}}$. Although there is good agreement between the model and transistor data, the above relation cannot be generalized, as a negative threshold voltage may also be obtained in some transistors measured in the literature.[8,16] In fact, in addition to hosting high concentrations of potential traps, the oxide surfaces are also particularly dipolar. Large dipoles can influence the threshold gate field of transistors built on this dielectric surface depending on the relative compositions (cations and anions) of the first oxide layers. Consequently, we believe that the oxide monopolar and dipolar effects add, in general, their strength to determine the threshold field. In our pentacene transistor, for instance, the plasma treatments on the oxide surface have essentially favored monopoles.

In conclusion, the ultrathin OFETs in this paper have successfully been used as model systems for describing the charge carrier propagation in pentacene layers and transport phenomena on the pentacene/oxide interface. The carrier transport is dominated by the first semiconducting layer where the plasma activated electron traps on the oxide interface induce equal amounts of residual holes, which determine the transistor transport characteristics and performance. Consequently, the oxide and pentacene layers should be treated together as a two-dimensional system. Finally, this work has stimulated us to reach the limit in ultrathin transistor fabrication and to understand the charge-transport processes at the interface between the pentacene and the $SiO_2$ gate dielectric so fundamental to applications.